\documentclass[conference]{IEEEtran}
\IEEEoverridecommandlockouts
\usepackage{cite}
\usepackage{amsmath,amssymb,amsfonts}
\usepackage{algorithmic}
\usepackage{graphicx}
\usepackage{textcomp}
\usepackage{xcolor}
\usepackage{float}
\usepackage[marginal]{footmisc}
\usepackage{setspace}
\begin{document}

\title{Metaverse Native Communication: A Blockchain and Spectrum Prospective}
\author{Hao Xu, Zihao Li, Zongyao Li, Xiaoshuai Zhang, Yao Sun, Lei Zhang
\thanks{
 		H. Xu, Zongyao Li, X. Zhang, Y. Sun and L. Zhang
 		are with the James Watt School of Engineering, University of Glasgow, UK; E-mail: 
 		\{H.Xu.2, L.Zongyao.1\}@research.gla.ac.uk;
 		\{Xiaoshuai.Zhang, Yao.Sun, Lei.Zhang\}@glasgow.ac.uk.
 		Zihao. Li is with CREATe Centre, School of Law, University of Glasgow, UK; E-mail: Z.Li.6@research.gla.ac.uk.
 	}
 	\thanks{Corresponding author: Lei Zhang.}
 	}


\maketitle

\begin{abstract}
Metaverse depicts a vista of constructing a virtual environment parallel to the real world so people can communicate with others and objects through digital entities. In the real world, communication relies on identities and addresses that are recognized by authorities, no matter the link is established via post, email, mobile phone, or landline. Metaverse, however, is different from the real world, which requires a single identity belongs to the individual. This identity can be an encrypted virtual address in the metaverse but no one can trace or verify it. In order to achieve such addresses to hide individuals in the metaverse, re-mapping the virtual address to the individual's identity and a specific spectrum to support the address-based communication for the metaverse are needed. Therefore, metaverse native or meta-native communications based on blockchain could be a promising solution to directly connect entities with their native encrypted addresses that gets rid of the existing network services based on IP, cellular, HTTP, etc. This paper proposes a vision of blockchain, encrypted address and address-based access model for all users, devices, services, etc. to contribute to the metaverse. Furthermore, the allocation architecture of a designated spectrum for the metaverse is proposed to remove the barrier to access to the metaverse/blockchain in response to the initiatives of metaverse and decentralized Internet.
\end{abstract}


\section{Introduction}

\IEEEPARstart{M}{ETAVERSE}, \footnotetext{This work has been submitted to the IEEE for possible publication.Copyright may be transferred without notice, after which this version may no longer be accessible. }the combination of the prefix ``meta'' with the word ``universe'', describes a hypothetical synthetic environment linked to the physical world \cite{Lee2021}. The word metaverse was first coined in a piece of speculative fiction named Snow Crash, written by Neal Stephenson in 1992. In this novel, Stephenson defines the metaverse as a massive virtual environment parallel to the physical world, in which users interact through digital avatars \cite{joshua2017information}. Metaverse is presenting an appealing world for cyber enthusiast and visionary people to take networking and society to the unknown edge. 

In recent, the consensus from multi mega-companies sets their goals to bring the metaverse to life by leveraging diverse techniques as the enablers such as virtual reality, artificial intelligence, human-computer interaction, internet of things, and blockchain \cite{meta2021,Lee2021}.

\subsection{Metaverse can be all about blockchain}
As a key building brick of metaverse, blockchain, stemming from the first decentralized digital cryptocurrency Bitcoin proposed by Satoshi Nakamoto \cite{Nakamoto2008}, is a distributed and immutable ledger that enables transparent transactions. Blockchain technology is a combination of modern cryptography, peer-to-peer network communication, distributed storage with consistency, and smart contracts to implement data exchange, process, and storage. A key benefit of blockchain is the decentralized consensus mechanism to realize anonymous and accountable transactions that may suitable for metaverse. In blockchain, when new blocks are verified, they can be linked to previous ones by cryptographic hash operation. As a consequence, blockchain can achieve tamper resistance by consensus mechanisms and public ledgers, i.e., it is computationally infeasible to distort the transactions that have been published in the blockchain. Therefore, such properties encourage blockchain to be considered as a fundamental component of metaverse, which is naturally distributed to address information exchange and storage, trade, access control, and so on \cite{van2021metaverse}. 

Metaverse was a concept, and it will still be a concept for a while, because the reality of metaverse is a collection of multiple technologies and services. However, the metaverse may only exist because of the universal record kept by a immutable record medium, e.g., blockchain, and that is the evidence of the existence of metaverse due to its immunity to changes and the resilience of decentralization. The strict definition on assets and ownerships on blockchain or other immutable records, may be the key to maintain a sustainable metaverse ecosystem. Meanwhile, preferences interpreted by other technologies (e.g., Virtual Reality (VR), Hepatic gloves, Holographic, etc.) to that extent, can be varied due to rendering configurations. 

Subtle differences of metaverse-ready virtual world experience and the existing virtual world experience are synchronous, persistent, independence in the world, and most importantly, the events have an integrated life-cycle in the metaverse, where they naturally start, develop and end.
In the future, metaverse service providers will allow users to interact using users' encrypted identities (which can be considered as the start of things in metaverse) and addresses instead of offering them account registration services for maximum user privacy and anonymity and end-to-end security. From that point, the metaverse has started forming with the blockchain as its first building block. In fact, the early age of metaverse has emerged with the encrypted identity (address) at the first instance of decentralized cryptocurrencies, the Bitcoin.

While the metaverse is likely to recognizes the blockchain's most important feature, non-fungible token (NFT), as the ownership identifier \cite{Wang2021}, it is also largely depending on the blockchain natural capability, where a record is transparent and immutable in the proper blockchain network. In our vision of metaverse, NFT might become the place for advertisements and the entrance to the service. The service provider can forge their NFTs for their content, and the content is accessible to the public. Such benefits make the NFT the gateway for advertised services.

\subsection{Accessibility of metaverse for all}
Leading by the blockchain, NFT and many distributed applications, Internet is pacing into its next revolution, where the decentralization banners are leading the way. However, in the current infrastructure, there is an inherent impossibility for the craved decentralization, as access to the network was never free to individuals \cite{TheNewYorkTimes2015}. To make the Internet a more decentralized place, the free and decentralized network access infrastructure such as a specific spectrum for decentralized networks should be on the list for Web 3.0 from public opinion \cite{TheNewYorkTimes2015}. On the other hand, decentralized autonomous organizations (DAO) with Blockchain and cryptocurrencies and NFT have made a huge fortune from the Internet, but their underlying networks have only paid a little cost of traffics to the whole network. As DAO, the stakeholder of blockchains, might be the core of decentralized network, hosting a web that is going to be built by all, it should consider the basic access right for all, and consider the maintenance of the network as its duty, in contradictory to current offerings. Hence, DAOs and Over-the-top (OTT) service providers are likely to be among the top of lists of eligible parties of affording the true cost of the network, the spectrum and the infrastructure that are going to make the Internet free for access.

\begin{figure*}[h]
	\centering
	\includegraphics[width=0.70\linewidth]{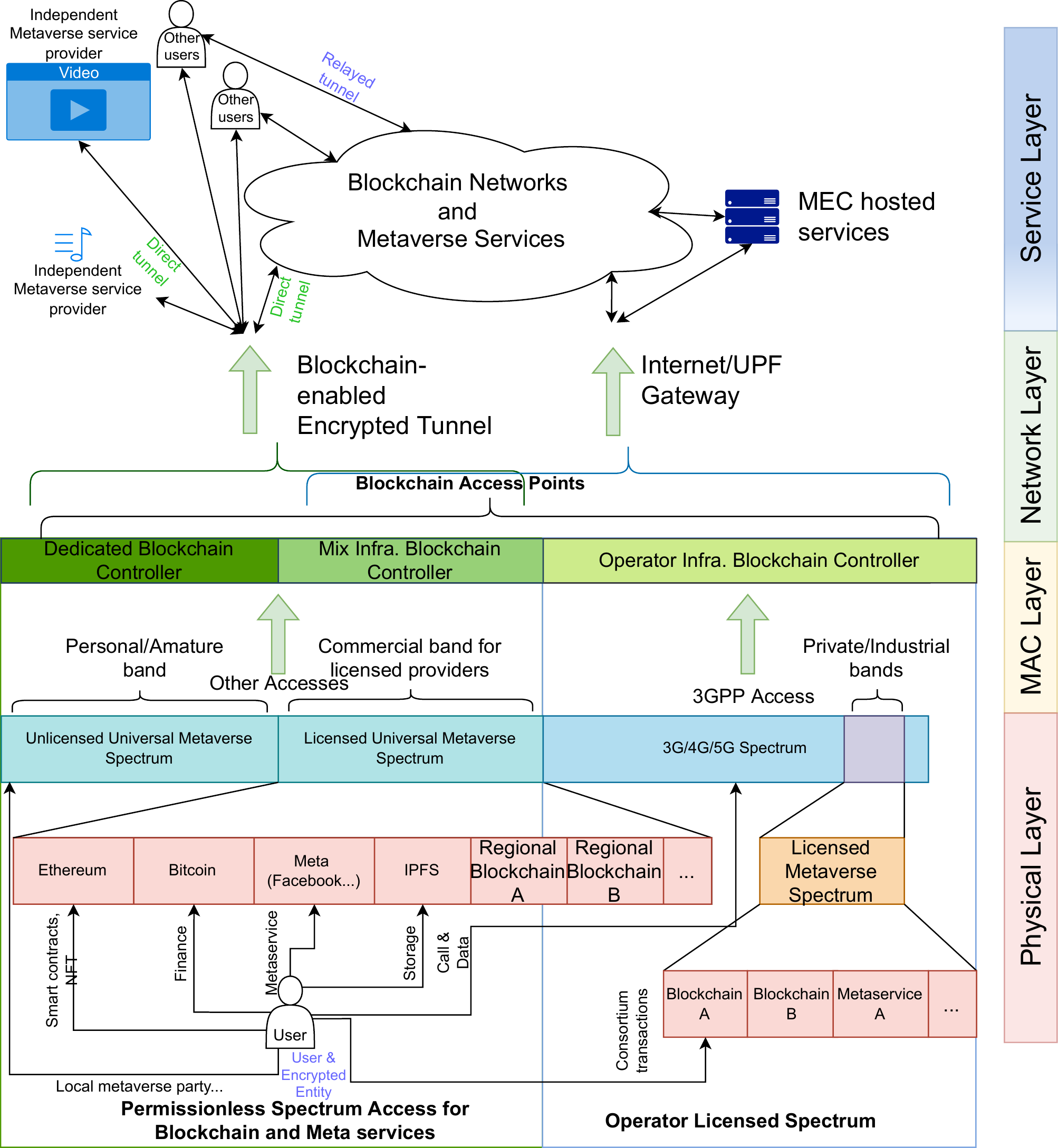}
	\caption{Metaverse Native Spectrum and Network Architecture}
	\label{fig:metaspectrum}
\end{figure*}

Since every entity may be known by its encrypted address in the metaverse, we envision that the communication should be optimized from the root level to ensure the reachability and security of every encrypted address with minimal third party involvement. Meanwhile, blockchain is natural to cryptocurrencies, and it generates huge profits while maintaining the network. Thereby, the benefactor, e.g., the DAO and any account holders of the cryptocurrency, has the reason to open up the access to the network. One feasible way is to let the blockchain and metaverse service provider own the dedicated service spectrum, making the access to the service free of access charge, as shown in Fig. \ref{fig:metaspectrum}, where the spectrum can be classed into unlicensed, licensed and operator owned, and a subclass under operator licensed spectrum for private blockchain/meta-spectrum. 
By introducing the metaverse spectrum, the service running on the spectrum can be totally metaverse native, and the access to such service can be assured by identity-based Service Level Agreement (SLA) or zero-rate policy. Further details on metaverse service spectrum are discussed in Section. \ref{sec:spectrum}.

\subsection{Motivations and Contributions}

Metaverse has a tight bind with blockchain, as it is naturally embracing encrypted identities, and the address associated with distributed ledger can be directly used for mutual authentication between address owners. The underneath stacks of identity management, authentications and transactions are the native ability of blockchain. Metaverse is also fundamentally related to the ownership and interactions on blockchain. Therefore, the only meta-information available in metaverse is the same encrypted address used for blockchain identity of all entities. No matter how the service changes, every entity can be tracked back to an encrypted address. However, employing blockchain in metaverse may incur high overhead since reaching a consensus in the blockchain requires massive communication and computing resources \cite{Zhang2021}. Therefore, communication and computing should be rooted deep from the metaverse and be metaverse-native. An architectural evolution of blockchain-enabled communication networks for the metaverse should be considered with an additional blockchain controller to the control plane.

Combined the above motivation, two key problems in utilizing blockchain to construct metaverse pop up: Who is paying for the massive communication and computing cost of metaverse? How do the governing bodies regulate the service on the metaverse, as it is completely decentralized and privacy-preserving? To this end, computing resource is also falling into the scope of metaverse resources, as an integrated assets within the metaverse. Moreover, besides the flexibility of accessing resource in the decentralized metaverse network, it also requires a way to quantify how much communication and computing resource used for a user or a service provider, thus to determine the cost and price of a metaverse service (not to be confused with free access to the metaverse and blockchain). Based on these, novel communication and computing resource managements are required in metaverse. 
	

In this paper, we explore how to support metaverse in wireless networks, where a novel Meta-native network has been proposed based on the Distributed Ledger Technology. In the Meta-native network, we first introduce an identity-based access model for all users, devices, and services with respect to encrypted addresses. Then, the management of both computing and communication resources has been discussed. More importantly, a concept of metaverse/blockchain dedicated service spectrum is defined for freeing the public blockchain access, ensuring the quality of blockchain services and the operation model between metaverse/blockchain service providers, operators and authorities. Finally, some challenges and future directions of the Meta-native network are given to further speed up the implementations.  

\section{Meta-native Communications and Computing}
We can assume that anything recorded on the blockchain or other recording medium can be considered as an entity in the metaverse, as they have their unique identity and address. By generating encrypted addresses, entities can be found on the blockchain, and users from the whole metaverse can acquire the connection information in a decentralized manner, where the identity is not issued by a central agency. Due to the decentralization of identity generation, trust among each other is considered minimal, and the connection between any entities should be mutually authenticated at first instance.

\subsection{Encrypted address is native to Metaverse}
The key to the reachability of encrypted address is to make every recorded encrypted address (i.e., the wallet address for blockchain transactions) connectable and directly via MAC layer or Network Layer (Layer 2 and 3) abstraction. It is critical to maintain the communication between each encrypted identity in a lower layer to remove further requirements of additional identities, in order to avoid identity leakage and injections. For instance, in Fig. \ref{fig:metaspectrum}, the service accessed by the user shall accept the user's encrypted address as its identity and provide services based on the agreement with the encrypted address, in comparison to the common IP addresses which are only used for the addressing purpose. 
The direct access can be achieved by building an encrypted tunnel overlay network, and one possible way is to make the blockchain, the medium of records, naturally routable, which is further detailed as a challenge in the later section.

Metaverse may only requires the user to have its wanted destination, without adding meta-external information, such as MAC address, International Mobile Equipment Identifier (IMEI), Subscriber ID, IP address, etc., these things should be hidden by the deep down network, and users should perform any actions without referring to any of them. It requires users to build a unified tunnel to contain the above routing and identity information, establishing secured end-to-end connections between the user and destination either by appointing the destination address or on ad-hoc basis if multiple entities are involved, as elabrated in Fig. \ref{fig:metaspectrum}. 

Therefore, Meta-native wireless communications should treat the encrypted address as its fundamental user information with support to implemented blockchain-enabled mutual authentication mechanism \cite{Xu2021beran} to connect all other devices in the network. In the case of mobile network, users should be able to choose whichever carrier for its metaverse data, if possible, even with the metaverse native operators. 

\begin{figure}[h]
	\centering
	\includegraphics[width=0.7\linewidth]{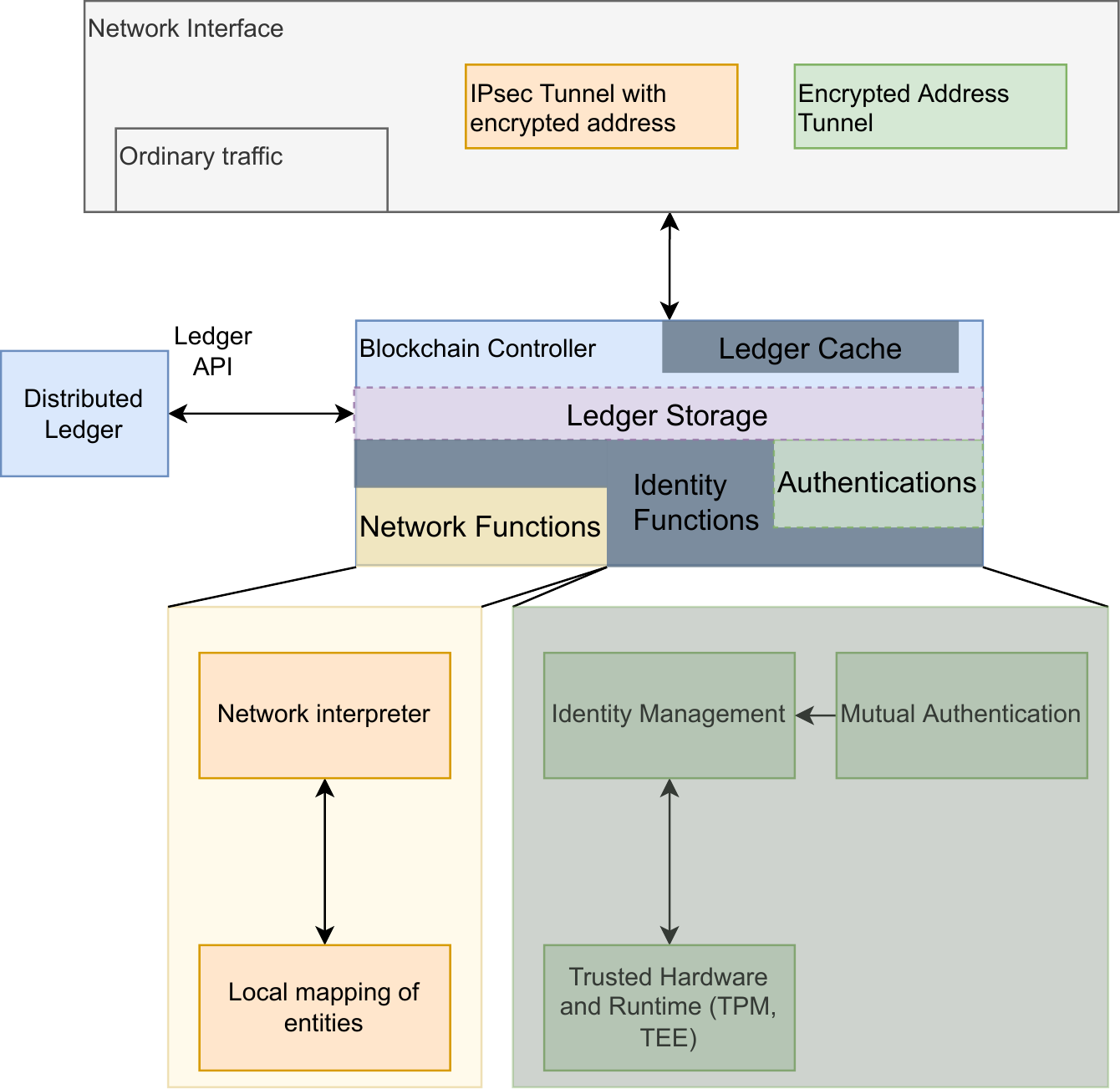}
	\caption{Blockchain controller for enabling encrypted address at Radio Access Network}
	\label{fig:metablockchaincontroller}
\end{figure}

\begin{figure}[h]
    \centering
    \includegraphics[width=0.70\linewidth]{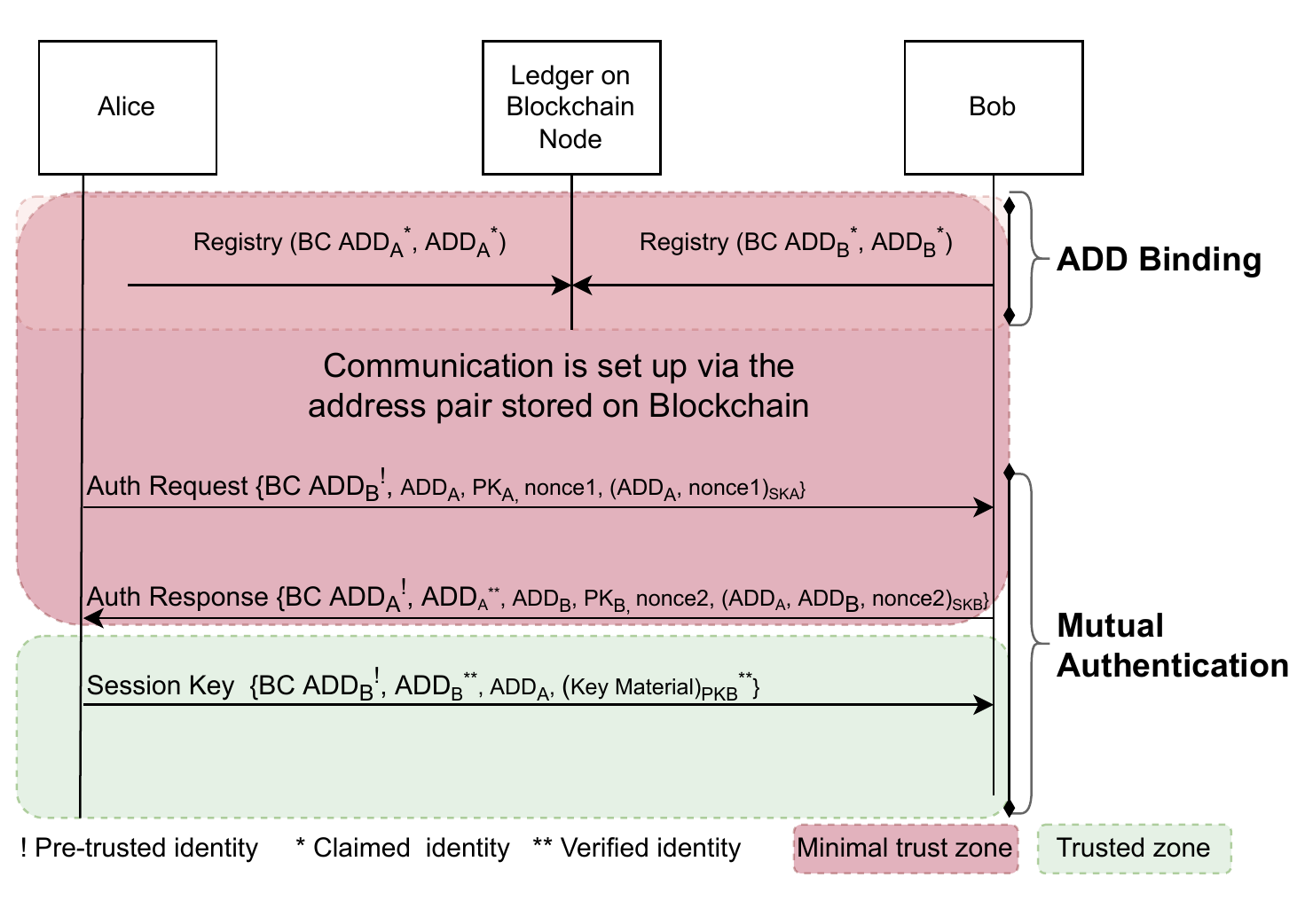}
    \caption{An example of blockchain-enabled mutual authentication \cite{Xu2021beran}}
    \label{fig:authentication}
\end{figure}

In order to gain the native support of encrypted addresses in the metaverse, the connection to the serving network shall be treated natively with the help of an embedded blockchain controller at the access network, as shown in Fig. \ref{fig:metablockchaincontroller}, where the controller acts as the agency for entities to interact with blockchain network and relay the information to entities who may not support blockchain access, concluded as Blockchain Access Points in Fig. \ref{fig:metaspectrum}. The controller is responsible for building the encrypted tunnel between two entities and forwarding the traffic as requested using its Network functions illustrated in Fig. \ref{fig:metablockchaincontroller} on the left. It also acts as a bridge to the conventional network and newly proposed encrypted network, with the network interpreter and translation of network addresses and encrypted addresses, as a part its identity functions. Meanwhile, the important feature, the authentication gateway for all entities shall be considered as integration at the access network level, and the authentication should be performed with the help of blockchain controller, indicated in Fig. \ref{fig:metablockchaincontroller} with supporting modules such as Trusted Execution Environment and Trusted Platform Module.
An example of blockchain-enabled mutual authentication for internal and external security is shown in Fig. \ref{fig:authentication}, where the blockchain is the medium of all entity addresses, and the user can initiate access requests with any entities in a decentralized, point-to-point manner using their encrypted addresses to request authentications. The details of referred authentication scheme can be found in \cite{Xu2021beran}.

\subsection{Computing resource and service sensing and direct access via communication}
Beyond the NFT as the real estate of metaverse, other resources should also be considered as the building blocks of metaverse and become the assets in metaverse based on encrypted addresses. Entity processors, for example, not only run their own systems, but they also process numerous requests, commands, transactions, calculations, and interactions with vast amounts of data in real-time. Therefore, computing resources are very key in the metaverse. Our terminals, such as VR devices, smartwatches, mobile phones, are extremely limited in terms of their processing capacity. In contrast, the computing power of servers or PCs is more ideal and it is possible to share computing resources in the metaverse. When we map processors into the metaverse, they become digital encrypted assets for metaverse users as well. Users can pay to use the asset as they would do the same in the real world. When their own computing resources are insufficient, they can use distributed communication means to send requests to the terminals with low processor occupancy around, which can maximize the utilization of computing resources in the metaverse. Therefore, distributed resource scheduling would be a very efficient way to utilize excess computing resource.

In metaverse, different communication methods can be used to provide computing services. Resource requests can be made through wireless direct access between devices. For example, Huawei has proposed a set of “Distributed Soft Bus” in Harmony operating system, which encapsulates all steps required for direct communication between devices into a set of protocols, including discovery, connection, authorization, authentication, etc. Devices equipped with HarmonyOS can interact through unimpeded wireless communication so that devices with weak computing capacity can benefit from those devices with strong computing power. Based on “Distributed Data Management Scheme” supported by HarmonyOS, it successfully solves the problem of allowing services to migrate on different terminals without interruption. Furthermore, service discovery can be easily obtained by broadcasting requests. If an entity capable of providing such a service receives the request, it may decide whether to establish a channel with the requester based on certain conditions. Meanwhile, the distributed ledger allows each entity to attach its own service list. Based on the consensus mechanism, users can access the latest service information in the blockchain and route to the specific entities when needed. In this way, the metaverse terminals can discover and connect with reachable local or global blockchain resources through data link layer and network layer with encrypted addresses and unified identification framework using wireless distributed soft bus and direct access protocol.

\subsection{Meta-native Wireless Access Network}
While we are re-defining the new paradigm of metaverse native communications and computing, wireless communications are important for local metaverse entities since most users connect with each other via mobile network, and in particular with nearby or remote metaverse related devices (VR/AR goggle, haptic gloves, computing hardware, etc.), which are meant to be accessed via owners/users encrypted identities, as the device is also an asset on the blockchain. Meanwhile, the privacy in Web 3.0 and metaverse may concern the users to avoid privacy-central communication network, as their privacy may be at risks from data breach and surveillance.
Thereby, one of the most important innovations for meta-native wireless communication is to enable the access with encrypted address, as illustrated in Fig. \ref{fig:metaspectrum} upon the network layer. With the state-of-the-art cellular network, users are bounded by the requirement of subscription of mobile network, not only for billing reasons, but also for the limitation of centralized architecture, as it needs pre-stored credentials to offer services to users and to discover users from its network, which is not feasible in the era of decentralization. The native interpretation of encrypted identities by wireless communication infrastructures can significantly improve the security, latency and scalability of local metaverse services while pushing the boundary of decentralization towards communication infrastructures. In addition to that, Mobile Edge Computing (MEC), on top right of Fig. \ref{fig:metaspectrum} can be integrated into such a paradigm without concerning identity and security challenges \cite{Kang2019}, or the need of core network to support to this extent.

\subsubsection*{Use case: Native encrypted address support on the control plane and user plane}
Once we have ruled out the native concept of metaverse communication, it is important to realize the access to the public radio access network with the private generated credentials, e.g., public key pairs and encrypted address (in blockchain flavor, shown in Fig. \ref{fig:authentication}), in fact, the credential is not only used to access the network, but also facilitating the connection in a permissionless manner. 

In the case of one user who wants to connect to a VR Head Mount Display (HMD), both the user's terminal device and the HMD are within the coverage of the mobile network. Taking into account that the HMD and the user have not been paired before, and they have no knowledge of each other, but the HMD has been declared on the blockchain with its encrypted address. 
Thanks to the declaration of HMD's encrypted address, the user is able to discover the HMD in the metaverse and initiates the connection. During the connection, their encrypted addresses will be used as authentication materials, exchanged using blockchain-enabled mutual authentication protocols, and establish a secured tunnel between them. 

In this use case, all information exchanged with blockchain network is carried by control plane, which plays an important role in signaling and access management. It is responsible for sending and receiving the access request from user/UE with its encrypted address, on par with its destination address. Note that, due to decentralized scheme of identity management, there is no central authority that keeps tracking on user's mobility and liveness, therefore, the connection is stateless unless the destination node replies. Once the base station receives the control plane signals with the encrypted address of destination, it looks up the address in the blockchain with the help of blockchain controller. The blockchain controller, as shown in Fig. \ref{fig:metablockchaincontroller}, on the one hand, helps the UE to find the route to the destination over the blockchain records. In the case of fully native encrypted address access scheme (e.g., routing is done over the blockchain), the blockchain controller will decide the flow of control plane message, if the entity is found within the same coverage of the base station or nearby communicable base stations, the flow of control plane message will be defined within the region, and the initial link between two encrypted entities will be established and mutual authentication over control plane starts.

Upon the successful initial connection, the two entities exchange a session key for further communication over user plane as indicated in Fig. \ref{fig:authentication}, and a point-to-point tunnel will also be established using the addresses and credential key materials provided. 
Using the established tunnel, user plane with encrypted addresses becomes the bearer of user data, and the service carried upon it may benefit from the mutual authentication for further identification. Above example shows the combination usage of control plane and user plane in the scope of wireless network for establishing point-to-point mutual authenticated encrypted communication.

\subsection{Blockchain and Metaverse Service Spectrum}\label{sec:spectrum}
In addition, blockchain as a general service is required by all entities and other services ubiquitously, and it is so important that comes to a necessity of having its dedicated serving spectrum and allocations for public available blockchain records, which can be known as Meta-spectrum, as shown in Fig. \ref{fig:metaspectrum}. Every user shall be able to use the dedicated or designated spectrum for wireless blockchain service without subscribing to operators, so the blockchain can always have the latest inputs from users and broadcast all latest blocks to users. Such Meta-spectrum shall be considered as the generic service spectrum for metaverse related public services and freely accessible. 

Operators, metaverse service providers, and DAO can bid for spectrum. Free access for all can mean many things, but a dedicated band for global/regional blockchain and metaverse service (as they are considered the universal building blocks and base for all metaverse concepts) is a good starting point, so global DAO, metaverse service companies can also bid for the global band of spectrum. On the other hand, regional spectrum bands can also be sold to local blockchain/metaverse service providers. The dedicated band can be used for public blockchain downlink and uplink transmission, so the access to the network can be totally independent, and free/tolled. Note that, the internal functions and structures of controllers in dedicated or shared spectrum are consistent, they are classified by their access to the certain resources in different domain. People may think the world of decentralization comes with free access to the network, and that is not necessarily true, but it is important to provide it as it was free. It is important to charge blockchain/P2P service via the spectrum if they are native to wireless.



\section{Challenges}
\subsection{Routing over blockchain and encrypted address}

As illustrated that metaverse system should be empowered by a blockchain-enabled distributed network, where the encrypted address is the only required information for meta-service provisioning. In this way, challenges of routing for data transmissions over these distributed nodes pop up due to the lack of IP address, which leads to the fact that classical routing protocols like IPv4/IPv6 cannot be applicable anymore. More importantly, the network topology in terms of node connectivity, link capacity, path reachability, etc. cannot be precisely obtained based on these encrypted addresses. In this case, even the two nodes with different encrypted addresses can interact with each other, it is not aware of how the nodes are connected (connected directly or via multi-hop). Therefore, it is quite challenging to design an efficient routing scheme for fully encrypted metaverse native communication. 

Besides the unawareness of network topology, the dynamics of service requests may also bring some extra difficulties on routing. Specifically, one physical metaverse node may have multiple encrypted addresses generated for different meta-service or logical entities. Considering the numerous  meta-service requests, the routing based on encrypted address should have high scalability. Moreover, not all the encrypted addresses are always in active mode due to the ending or sleeping of the service. Hence, a dynamic routing scheme for data forwarding/transmission should be required with respect to the transition of active and inactive mode.

Meanwhile, there is no dedicated entity (like routers in traditional computer networks) in the meta-native communication system taking care of routing function. Hence, it is rather challenging to determine who and how to update/analyze/store routing-related data thus achieving a low-latency and high reliability of data transmission. Furthermore, there is one practical issue that some users/nodes may have low willingness to help others forward data.
A potential solution is to exploit incentive schemes from blockchain system to motivate users to participate in routing-related activities. However, a careful design of the incentive scheme is required in this case. 

In addition, the encrypted identity framework is also leading to concerns from regulatory perspectives, i.e., how should law regulate such encrypted network. Compared with the traditional online infrastructure, the blockchain-native metaverse does not allow third parties to access the content of services and all information is encrypted and encapsulated into the point-to-point encrypted tunnel, which means that legal enforcement is difficult to interfere. Such encrypted network is likely to be a shroud for criminal activity and therefore an obstacle to law enforcement \cite{EuropeanPoliceOffice2016}. Efforts to combat crime and illicit behavior in metaverse are weakened or broken by such encryption infrastructure. It has been pointed out that encryption is a key threat and serious impediment to the detection, investigation, and prosecution of such criminal activity. However, such dilemma is not only confronted by this project. An alternative solution could be that more legal elements should be taken into account in design stage to achieve regulation-by-design, which could avoid conflicts in advance. Metaverse as a brand-new sphere also needs to establish its own legal and ethical framework to regulate users’ behaviors. Therefore, further research needs from both legal and engineering perspectives to explore the regulatory rules in metaverse.

\subsection{Spectrum allocation and wireless capacity}
Because spectrum is a limited national resource and most countries consider it as an exclusive property of state, it leads to another potential challenge that how to convince governing bodies for designated spectrum allocation. 
As a result, a spectrum market could be introduced to trade the spectrum in the secondary market. Such a spectrum property right could assure the ownership of spectrum and lead to more efficient spectrum usage, since spectrum owners are highly likely to economize their resources \cite{Berry2010} for profits.

While trading the spectrum is boosting the capacity of wireless metaverse and blockchain, the spectrum is still regarded as a limited resource, both in licensed and unlicensed range. The emerging metaverse services, e.g., Mixed Reality (MR), Augmented Reality (AR), etc., requires huge bandwidth that may be not feasible for large scale coverage, hence the service providers are required to ensure the capacity of wireless access network on the spot. With commercialized spectrum exchange, the service providers have more choice on their expansion strategy, choosing between operator managed network or self-managed network.  

\subsection{Post-quantum Metaverse}
Encrypted identity and address face challenges from quantum computing, in particular, the public key scheme adopted in the vision. An urgent need to find post-quantum public key algorithm is eminent \cite{del2018lattice}, as the discrete logarithm problem is no longer a challenging problem in the realm of quantum computation with \cite{Shor1997}. Many quantum proof algorithms have emerged with the introduction of Lattice-based cryptography \cite{del2018lattice}, which is seen as the next generation of public key basis, it helps public key to prevent quantum supremacy against public key based encrypted identity, the vision presented in this paper shall be secure thereafter.

\section{Conclusion}
In this paper, we illustrate the insight of how the blockchain can contribute to the metaverse as the blockchain is native to the metaverse concept. Blockchain gives the metaverse a solid course towards reality, and powers up the identity framework, service ecosystems for the metaverse, where the encrypted address can be regarded as the entry point for all entities in the metaverse. Furthermore, we discuss the necessity to promote specific spectrums for blockchain and metaverse to enable free access to services and devices for users, access management for business use cases, and the ecological innovation on the network maintenance in a highly decentralized manner.

\bibliographystyle{IEEEtran}
\bibliography{IEEEabrv,ref}

\end{document}